\documentstyle[multicol,aps,psfig]{revtex}
\renewcommand{\narrowtext}{\begin{multicols}{2}
\global\columnwidth20.5pc\noindent}
\renewcommand{\widetext}{\end{multicols}
\global\columnwidth42.5pc}
\begin{document}
\draft
\preprint{February 1999}
\title{Spin flop in one-dimensional quantum antiferromagnets}
\author{T$\hat{\mbox o}$ru Sakai}
\address
{Faculty of Science, Himeji Institute of Technology,
 Ako, Hyogo 678-1297, Japan}
\date{February 1999}
\maketitle
\begin{abstract}
The field-induced transition corresponding to the spin flop 
in the quantum antiferromagnetic chains in the presence of 
the Ising-like single-ion anisotropy $D$ $(<0$)  
is studied by the finite-cluster analysis.  
It is found that for $S\geq {3\over 2}$ 
a first-order metamagnetic transition occurs even in one dimension 
for small negative $D$ except for the Haldane phase, 
while there exist two second-order transitions and  
an incommensurate phase appears between them 
for intermediate $D$ 
, as already found for $S=1$. 
We also discuss on 
some experiments of CsNiCl$_3$ related with the present 
work.  

\end{abstract}
\narrowtext

Magnetization measurements of antiferromagnets 
detect various macroscopic quantum effects.
One of interesting examples appear in the spin flopping (SF) 
process of low-dimensional materials.  
SF is one of field-induced metamagnetic phase transitions
of anisotropic Heisenberg antiferromagnets.\cite{neel} 
In terms of the classical spin systems, 
the transition brings about an abrupt change in the direction 
of the N\'eel order from parallel to perpendicular with respect to 
the easy axis under the applied field $H$ along the axis 
and it is a first-order transition with a jump in the magnetization $m$
as a function of $H$.  
In quantum systems the jump generally shrinks due to quantum fluctuation 
and such an effect becomes larger in lower dimensions and for smaller
$S$.  
For $S={1\over 2}$ antiferromagnets with the Ising-like 
anisotropic exchange coupling,   
the magnetization jump was revealed to survive down to two dimension
by the recent numerical analysis\cite{kohno}, 
while the exact solution\cite{yang} by the Bethe ansatz suggested  
that it changes into a second-order transition descibed by the 
asymptotic behavior 
\begin{equation}
   m\sim (H-H_c)^{1/\delta }
   \label{hc}
\end{equation}
where $\delta =2$ in one dimension (1D). 
Thus the large quantum fluctuation in 1D seems to change the order 
of the transition in the system.  

In view of experimental realization, one of the most important 
anisotropy is the single-ion anisotropy (SIA) described by 
$D\sum _j(S^z_j)^2$ which is relevant only for $S\geq 1$. 
In the classical limit in any dimensions 
the negative $D$ makes the N\'eel order oriented along $z$-axis and 
the first-order SF transition is induced by the applied 
field parallel to the axis. 
The recent numerical analysis\cite{sakai} 
suggested that instead of SF two successive  
second-order transitions occur in 1D $S=1$ antiferromagnet for 
small negative $D$ except for the Haldane phase.\cite{haldane} 
This is because the magnetic phase of the system consists of the two 
gapless Luttinger liquid phases 
with different elementary excitations created by 
$(S^+_j)^2$ and $S^+_j$, respectively.   
The first transition, described by the form (\ref{hc}) with $\delta =2$,  
occurs from the N\'eel state to the first 
massless phase characterized by the dominant spin correlation function 
$\langle S^z_0S^z_r \rangle \sim \cos (2k_Fr)r^{-\eta _z}$,  
where $2k_F=2\pi m$, and 
the next transition with $\delta =1$ 
leads to the second massless phase where 
$\langle S^+_0S^-_r \rangle \sim (-1)^rr^{-\eta}$. 
The behavior of the spin correlation function suggests that 
the first massless phase, which lies 
between the two transitions, has no
correspondence in the classical systems.  
It implies that the appearance of the phase is an intrinsic 
quantum effect. 
Is such a strongly quantized behavior general in 1D, or peculiar  
to $S=1$? 
The question encourages us to investigate 1D larger-$S$ systems. 
In this paper, 
using the finite-cluster analysis, 
we study the magnetization process of the 
1D $S={3\over 2}$ and 2 antiferromagnets in the presence of the 
Ising-like SIA, in comparison with the case of $S=1$.

The 1D Heisenberg antiferromagnet with SIA  
under magnetic field parallel to the easy axis of the N\'eel order  
is described by the Hamiltonian 
\begin{eqnarray}
\label{ham}
&{\cal H}&={\cal H}_0+{\cal H}_Z, \nonumber \\
&{\cal H}_0& = \sum _j {\bf S}_j \cdot {\bf S}_{j+1}
+D\sum _j(S_j^z)^2, \\
&{\cal H}_Z& =-H\sum _j S_j^z, \nonumber
\end{eqnarray}
where ${\bf S}_j^2=S(S+1)$ and the periodic boundary condition are 
applied. We restrict us on the Ising-like anisotropy $D <0$ to
investigate the transition corresponding to SF. 
For $L$-site systems,
the lowest energy of ${\cal H}_0$ in the subspace where
the eigenvalue of $\sum _j S_j^z$ is $M$
(the bulk magnetization is $m=M/L$) and
the momentum is $k$, is denoted as $E_k(L,M)$.
In addition we define $E(L,M)$ as the lowest one among $E_k(L,M)$'s.
Using Lanczos algorithm, we calculated $E_k(L,M)$
for even-site systems up to $L=20$, 14 and 12 for $S=1$, ${3\over 2}$
and 2, respectively. 

The magnetic ground state (GS) 
for $S>1$ is divided into two massless phases
\cite{schulz} in the phase diagram on the $m/S$-$D$ ($D<0$) plane 
as well as $S=1$. 
The argument on the two phases for $S=1$ developed in the Ref.[4]
is easily generalized for arbitrary $S$ as follows:  
(i)Large negative $D$ phase; the gapless excitation is created by 
$(S^+_j)^{2S}$ and the soft mode has the momentum 
$2k_F=(1-{m\over S})\pi$. 
(ii)Small negative $D$ phase; the $S_j^+$ excitation is gapless and 
the soft mode has $2k'_F=2mS\pi$. 
The dominant spin correlation functions are 
(i)$\langle S_0^zS_r^z\rangle \sim \cos (2k_Fr)r^{-\eta_z}$ and 
(ii)$\langle S_0^+S_r^-\rangle \sim (-1)^r r^{-\eta}$, 
respectively.  
Thus in quasi-1D the coherent interchain interaction is expected to 
result in the long-range (i)incommensurate SDW along the $z$-axis and 
(ii)N\'eel order in the $xy$-plane (called the canted N\'eel order 
in some texts), respectively.  
In this paper we call the two massless phases (i)`$z$-SDW' and
(ii)`$xy$-N\'eel' phases to emphasize the long-range order 
expected to be observed in real quasi-1D systems, 
whereas pure 1D systems have some power-low decays in the associated  
spin correlation functions. 

Consider the two energy gaps defined as 
\begin{eqnarray}
\label{gap1}
&\Delta _1&=E(L,M+1)+E(L,M-1)-2E(L,M),  \nonumber \\
&\Delta_{2k_F}&=E_{2k_F}(L,M)-E(L,M).
\end{eqnarray}
$\Delta _1$ is the sum of the gaps corresponding to $\delta M=1$ and 
$-1$ for the total Hamiltonian $\cal H$. 
Since $\Delta _1$ ($\Delta _{2k_F}$) is open (gapless) in the   
$z$-SDW phase, while gapless (open) in the $xy$-N\'eel phase, 
the crossing point of the two gaps for fixed $m$ is a good estimation 
of the phase boundary. 
Since only states with $M=2Sn$ $(n=0,1,2,\cdots)$ contributes 
to the magnetization process 
in the $z$-SDW phase, our analysis is focused on such
states. 
The crossing points of the finite clusters with various magnetizations 
are plotted on the $m/S$-$D$ plane as open symbols 
in Figs. 1 (a)$S={3\over 2}$ and (b)$S=2$.     
Little size dependence of the curves justifies that the fitted solid
lines are the phase boundary of the bulk systems. 
$D_{c0}$ denotes the boundary between the N\'eel ordered and disordered
phases in the nonmagnetic GS. 
It corresponds to the boundary of the Haldane phase for integer $S$, 
while $D_{c0}=0$ for half-odd integer $S$. 
Using the phenomenological renormalization\cite{prg}, 
$D_{c0}$ of the $S=2$ system is estimated as $D_{c0}=-0.001\pm 0.001$. 
The phase boundary at $m/S=1$ denotes $D_{c2}$. 
It was determined as the crossing point of $E(L,SL)-E(L,SL-1)$ and 
$E(L,SL)-E(L,SL-2S)$ which is almost independent of $L$. 
The results are $D_{c2}=-1.31$ and $-1.25$ for $S={3\over 2}$ and 2, 
respectively.  
The $m/S$-$D$ phase diagrams suggest that 
besides the two ordinary critical fields $H_{c1}$ and $H_{c2}$,  
which are the starting and saturation points of the magnetization, 
respectively, the magnetization process has 
an intermediate critical field $H_{c3}$ ($H_{c1}<H_{c3}<H_{c2}$)  
corresponding to the boundary of the two massless phases 
for $D_{c0}>D>D_{c2}$. 
According to the argument for $S=1$, 
the phase transitions at the three critical points  
are all second-order and have the asymptotic forms 
\begin{eqnarray}
\label{asym1}
m\sim (H-H_{c1})^{1/2},  
\end{eqnarray}
\begin{eqnarray}
\label{asym3}
m-m_{c3}\sim |H-H_{c3}|,
\end{eqnarray}
\begin{eqnarray}
\label{asym2}
S-m\sim (H_{c2}-H)^{1/2},
\end{eqnarray}
respectively, 
where $m_{c3}$ is the magnetization corresponding to the boundary  
between the two massless phases.   
In the classical limit the first-order SF transition occurs
and $m$ jumps from 0 to $[-D/(D+2)]^{1/2}S$ (long-dashed
lines in Figs. 1 (a) and (b)) for $0>D>-1$, 
while 0 to $S$ for $D<-1$. 
The phase diagrams including only the two massless phases 
might lead to the conclusion that 
even for larger $S$, as well as $S=1$, 
two second-order transitions occur instead of SF 
for $D_{c0}>D>D_{c2}$ 
and the phase boundary between the two massless phases 
will go to the classical line 
as $S$ increases towards $\infty$.  
\begin{figure}
\mbox{\psfig{figure=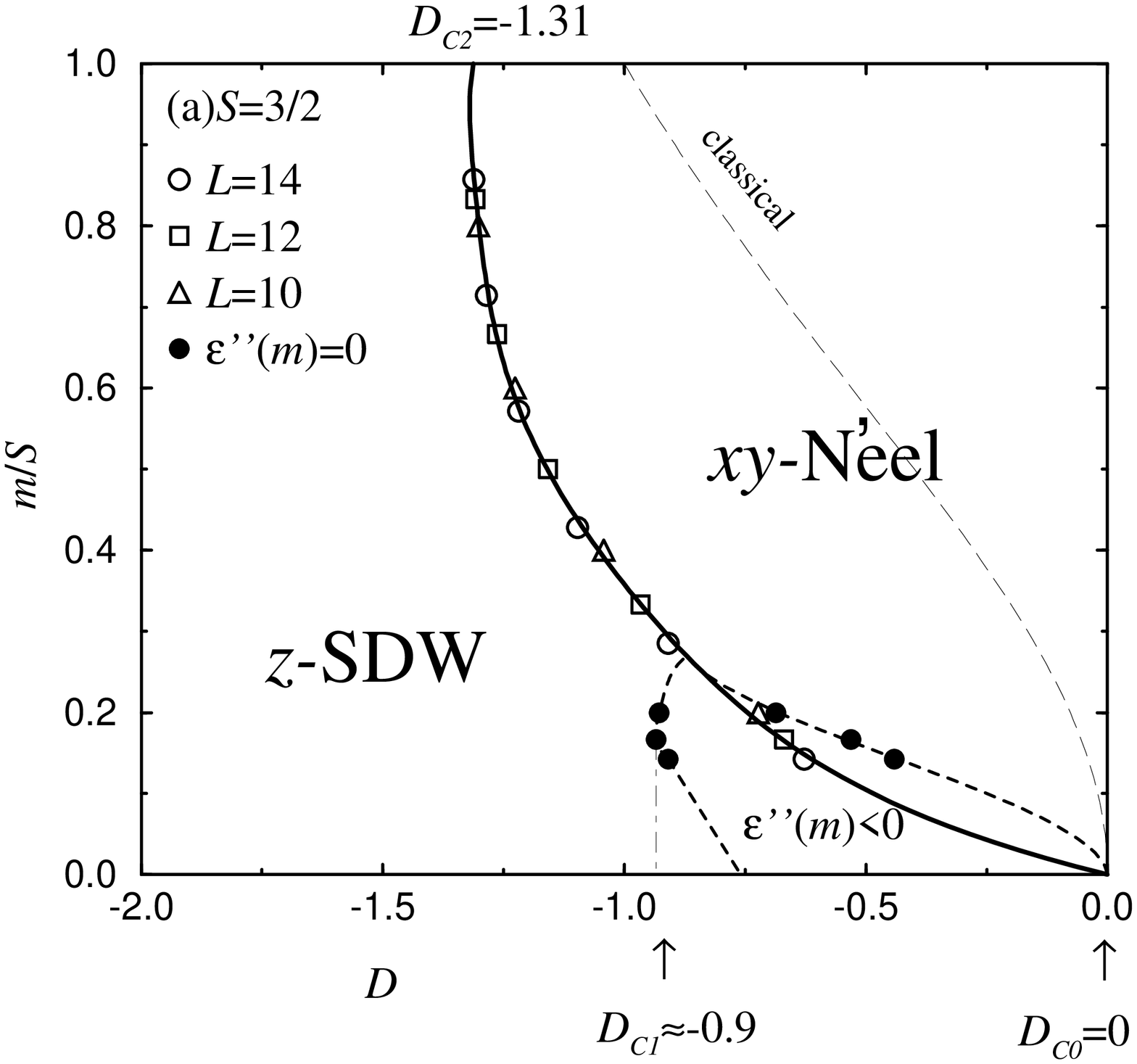,width=85mm,angle=-90}}
\vskip 0.1pt
\mbox{\psfig{figure=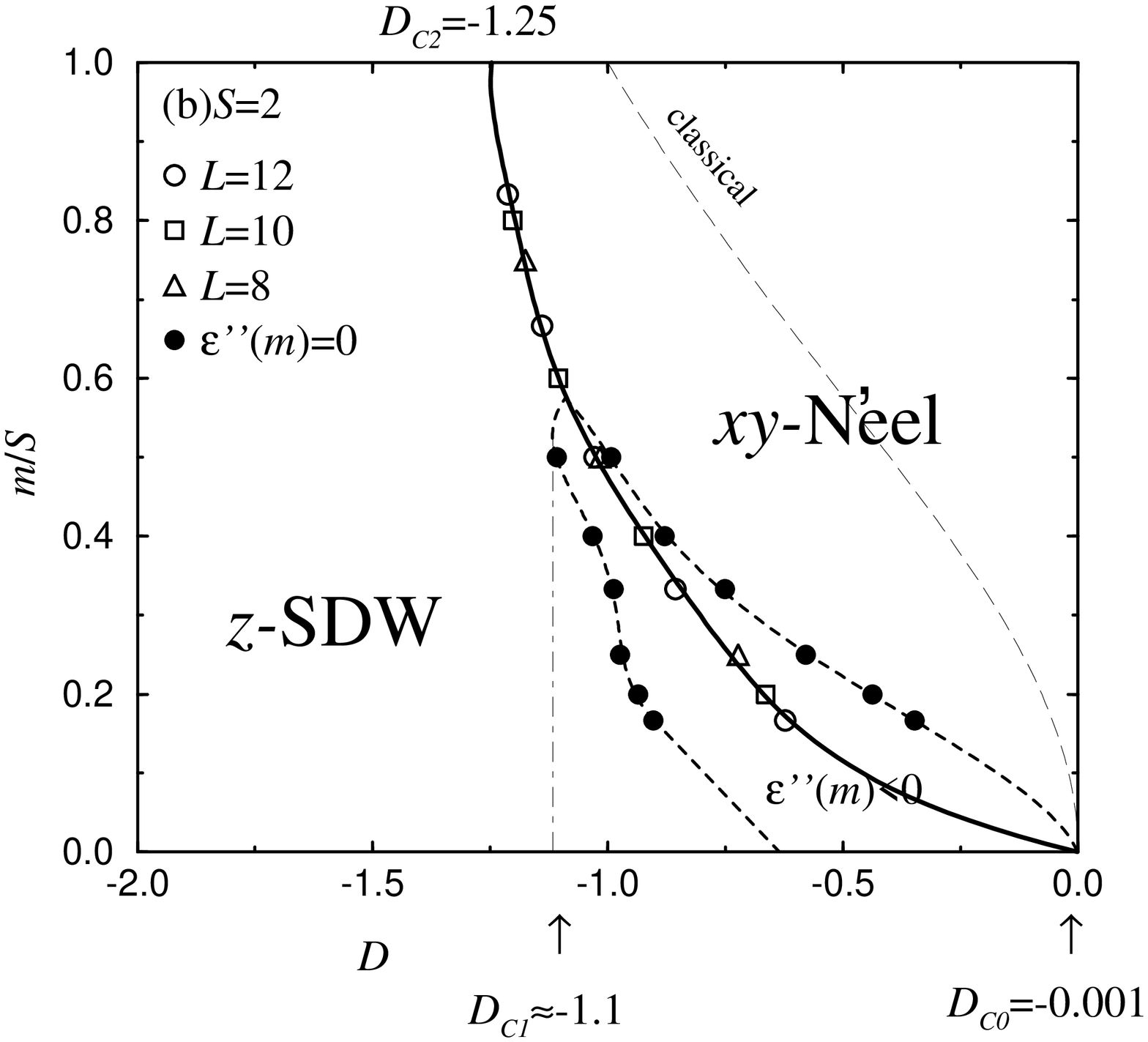,width=85mm,angle=-90}}
\vskip 5mm
\caption{
$m/S$-$D$ phase diagrams for (a)$S={3\over 2}$ and (b)$S=2$. 
Open symbols are the crossing point of the two gaps $\Delta _1$ 
and $\Delta _{2k_F}$ for finite systems. The fitted solid curves are 
the boundaries between the $z$-SDW and $xy$-N\'eel phases. 
Solid circles are the points with $\epsilon ''(m)=0$ and 
the curvature $\epsilon ''(m)$ is negative in 
the regions surrounded them. (Dashed curves are guides for the eyes.)
}
\label{fig1}
\end{figure}
\vskip 5mm

In the following analysis, however, 
it will be found that the first-order 
SF transition can occur for $S\geq {3\over 2}$. 
The lowest eigenvalue of the Hamiltonian ${\cal H}_0$ 
for $0<m<S$ per site 
in the thermodynamic limit, denoted as $\epsilon (m)$,  
can be estimated using the form\cite{cft}
\begin{eqnarray}
\label{gse}
{1\over L}E(L,M) \sim \epsilon (m) +C(m) {1\over {L^2}},
\end{eqnarray}  
where the second term is the size correction. 
As the GS magnetization curve is derived from 
$H=\epsilon '(m)$, 
a necessary condition of the continuity  
of the curve at $m$ 
is $\epsilon ''(m)\geq 0$. 
Since only $M=2Sn$ is available in the $z$-SDW phase, 
we investigate the condition using the form
\begin{eqnarray}
\label{deriv1}
&R&(L,M) \equiv  \nonumber \\ 
&L&[E(L,M+2S)+E(L,M-2S)-2E(L,M)]/(2S)^2 \nonumber \\
&\sim & \epsilon ''(m)+O({1\over L^2}), 
\end{eqnarray}  
derived from (\ref{gse}). 
$R(L,2S)$ is plotted vs $D$
for $S=1$, ${3\over 2}$ and 2 in Figs. 2 (a), (b) and (c), respectively. 
They indicate the existence of 
the region with $R(L,2S)<0$ for $S={3\over 2}$ and 2, 
and it tends to extend towards $D=0$ with increasing $L$, 
in contrast with the case of $S=1$ where $R(L,2S)$ is always positive.   
Neglecting the size correction in (\ref{deriv1}) 
and taking $R(L,M)$ for the 
largest $L$ available for each $m$ as $\epsilon ''(m)$, 
the points with $\epsilon ''(m)=0$ 
are plotted as solid circles in Figs. 1 (a) and (b). 
In the area surrounded by them $\epsilon ''(m)$ is negative.  
The fitted dashed lines are guides for the eyes drawn assuming 
that the negative $\epsilon ''(m)$ region continues to $D_{c0}$. 
Only for $S=1$,  
$\epsilon ''(m)\geq 0$ is satisfied everywhere 
on the $m$-$D$ plane within the same analysis up to $L=20$. 
For $S={3\over 2}$ and 2 
the existence of the area with $\epsilon ''(m)<0$ implies that 
the first-order SF transition occurs and  
$m$ jumps over the region.
Figs. 1 (a) and (b) suggest that SF can appear only in the 
small negative $D$ region. This is because 
the large negative $D$ strongly quantizes the system 
and makes it equivalent 
to the $S={1\over 2}$ $XXZ$ model which is proved to have no first-order
transition.  
Obviously the negative $\epsilon ''(m)$ region is larger for $S={3\over
2}$ than $S=2$, which is consistent with the assumption that 
smaller-$S$ systems have larger quantum fluctuation suppressing  
the magnetization jump. 
The large negative $D$ border of the SF region, 
denoted as $D_{c1}$,  
is roughly estimated
from the boundary of the negative  $\epsilon ''(m)$ region for the
finite systems as shown in Figs. 1 (a) and (b). 
The results are $D_{c1}\simeq -0.9$ and $-1.1$ for $S={3\over 2}$ and 
2, respectively. 
The size correction neglected on the estimation of $D_{c1}$ is not 
supposed to be so large as to change the following qualitative 
argument, because Figs. 2 (b) and (c) suggest that the left border 
of the region with $R(L,M)<0$ has no
significant $L$ dependence in comparison with the right border. 
For $D_{c0}>D>D_{c1}$ the magnetization process is expected to have 
a jump from $m=0$ to the $xy$-N\'eel phase. 
Figs. 1 (a) and (b) might suggest the possibility of the jump from  
nonzero $m$ near $D_{c1}$, but it is difficult to determine 
the point of the jump within the present small cluster analysis
because the first order transition is 
ruled by the Maxwell construction, 
which cannot be applied until a more precise curve 
of $\epsilon (m)$ is given. 
\begin{figure}
\mbox{\psfig{figure=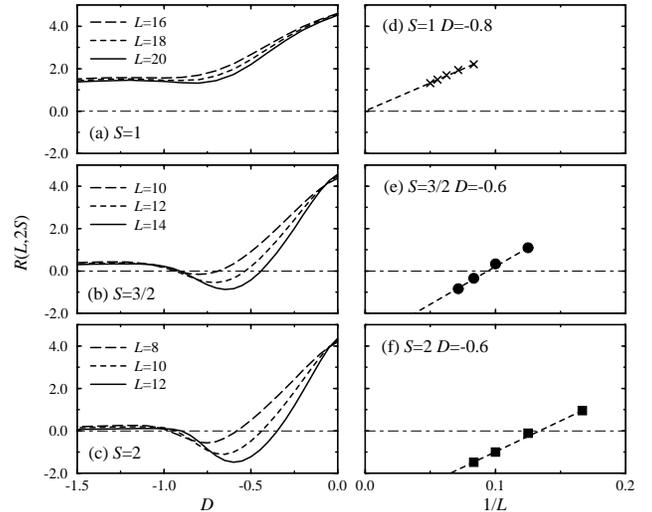,width=85mm,angle=-90}}
\vskip 5mm
\caption{
$R(L,2S)$ vs $D$ of the finite systems for (a)$S=1$, 
(b)$S={3\over 2}$ and (c)$S=2$. 
$R(L,2S)$ vs $1/L$ with fixed $D$; 
(d)$S=1$ $D=-0.8$, (e)$S={3\over 2}$ $D=-0.6$ and 
(f)$S=2$ $D=-0.6$. 
These values of $D$ are around the minima (local minimum for $S=1$) 
of the largest-$L$ curves in (a), (b) and (c). 
Dashed lines are based on the least square 
fitting. 
}
\label{fig2}
\end{figure}
\vskip 5mm

To convince of the existence of SF  
for $S={3\over 2}$ and 2 
in the thermodynamic limit, 
we investigate the behavior of $\epsilon (m)$ around the limit 
$m\rightarrow 0+$. 
Assuming that the size correction of $\epsilon (m)$ is at most 
$O(1/L^2)$ even in the massive phase ($m=0$), 
the size dependence of $R(L,2S)$ is expanded around $m=0$ as  
\begin{eqnarray}
\label{deriv2}
R(L,2S)  
\sim \epsilon ''(0)+ (2S)\epsilon'''(0){1\over L}+  O({1\over L^2}).
\end{eqnarray}  
Around the minimum (local minimum for $S=1$) 
of the $R(L,2S)$-$D$ curve in Figs. 2 (a-c)  
for the largest $L$ ($D=-0.8$ for $S=1$ and $D=-0.6$ for 
$S={3\over 2}$ and 2), the size dependence of $R(L,2S)$ is investigated
by the plot of $R(L,2S)$ vs $1/L$ in Figs. 2 (d)$S=1$, (e)$S={3\over 2}$ 
and (f)$S=2$,  
where dashes lines are the results 
from the least square fitting. 
Considering the form (\ref{deriv2}), 
the plots for $S={3\over 2}$ and 2 both suggest $\epsilon ''(0)<0$ and 
$\epsilon '''(0)>0$ which is a sufficient condition of the 
appearance of the magnetization jump from $m=0$ at some critical field
in the thermodynamic limit. 
On the other hand, 
the plot for $S=1$ implies  $\epsilon ''(0)=0$ and 
$\epsilon '''(0)>0$ which is consistent with the second-order 
transition at $H_{c1}$ described by the asymptotic form (\ref{asym1}).   
These results also support the existence of SF 
for $S\geq {3\over 2}$ in the thermodynamic limit.  
Since the quantum fluctuation becomes smaller for larger $S$, 
the SF region should become  
larger with increasing $S$ 
and in the limit $S\rightarrow \infty$ the region 
would occupy all the region at the left 
hand side of the classical SF line in the $m/S$-$D$ 
phase diagrams in Figs. 1 (a) and (b). 
\begin{figure}
\mbox{\psfig{figure=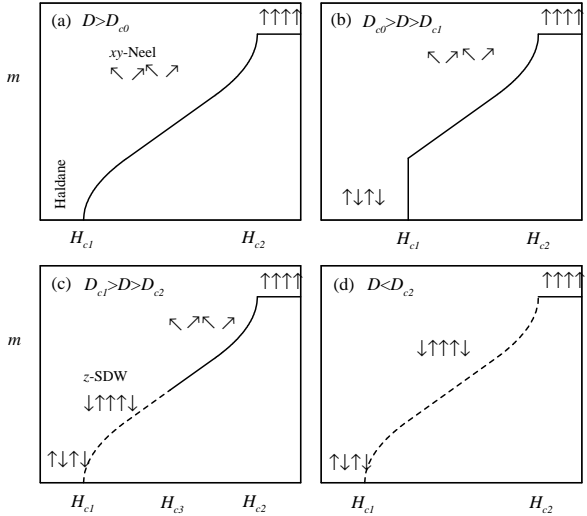,width=85mm,angle=-90}}
\vskip 5mm
\caption{
Schematic magnetization curves of four possible cases. 
Solid and dashed curves signify the $xy$-N\'eel and 
$z$-SDW phases, respectively, except for massive regions.  
}
\label{fig3}
\end{figure}
\vskip 5mm
  
Based on the $m/S$-$D$ phase diagrams  
in Figs. 1 (a) and (b), 
we summarize all possible cases of the magnetization process of 
1D Heisenberg antiferromagnet with arbitrary $S$ ($>{1\over 2}$) 
in the presence of the Ising-like SIA.  
In the following discussion 
the critical fields $H_{c1}$ and $H_{c2}$ always stand for  
the starting and saturation points of the magnetization.  
The schematic magnetization curves are shown in Figs. 3 (a), (b), 
(c) and (d), and their features are listed as follows: 
(a)$D>D_{c0}$ (only for integer $S$); 
the second-order transition with the asymptotic form (\ref{asym1}) 
from the Haldane to $xy$-N\'eel phases occurs at $H_{c1}$. 
(b)$D_{c0}>D>D_{c1}$; 
the first-order SF transition occurs at $H_{c1}$ from the 
N\'eel to $xy$-N\'eel phase. 
The present analysis leads to the absence of the case 
only for $S=1$. 
(c)$D_{c1}>D>D_{c2}$; 
the second-order transition from the N\'eel to $z$-SDW phases at 
$H_{c1}$ described by the form (\ref{asym1}) is followed by 
another second-order transition to the $xy$-N\'eel phase at $H_{c3}$ 
with the form (\ref{asym3}). 
(d)$D<D_{c2}$; 
the second-order transition from the N\'eel to $z$-SDW phases 
and the latter phase continues until the saturation. 
The last case (d) in the large negative $D$ region 
is equivalent to the Ising-like $S={1\over 2}$ $XXZ$ chain. 
In any case the transition at the saturation point $H_{c2}$ is  
a second-order one described by the critical behavior (\ref{asym2}). 
As $S$ increases towards the classical limit, 
the boundaries $D_{c1}$ and $D_{c2}$ will go to the same limit 1 
and the case (c) will disappear in the limit.  
For $D<D_{c2}$ $H_{c1}$ and $H_{c2}$ will become close to each other 
and coincide in the limit $S\rightarrow \infty$. 

The most interesting case is (c) in Fig. 3 
where the two second-order transitions
take place and the $z$-SDW phase appears between them. 
The most realistic system where such a magnetization process can be 
realized is the $S=1$ chain.  
One of good candidates to observe the process is expected to be 
CsNiCl$_3$, a quasi-1D $S=1$ antiferromagnet 
with the Ising-like SIA, which has the N\'eel order
at low temperatures.\cite{csnicl3}  
The high-field magnetization measurements\cite{nojiri,katori} 
on CsNiCl$_3$ 
indicated a SF-like transition. 
However, the experimental magnetization curve 
also looks like a second-order transition described by 
the critical behavior (\ref{asym1}), 
or two successive transitions accompanied by an intermediate phase.
\cite{ajiro} 
In addition the NMR measurement\cite{inagaki} on CsNiCl$_3$ 
indicated two spin structures 
oriented in different directions around the transition. 
It may imply that another transition from the $z$-SDW to $xy$-N\'eel 
phases occurs soon after the first one. 
It would be difficult to detect the latter transition at $H_{c3}$ in 
the measured magnetization curve, because the transition brings about 
no significant anomalous behavior in the curve.
\cite{sakai} 
As well as NMR, 
the ESR or neutron scattering  
would be a better method to detect the change in the direction 
of the dominant spin correlation or the long-range order 
from parallel to perpendicular with respect to $H$ 
on the transition at $H_{c3}$. 
It would be interesting to observe the $z$-SDW phase which is an 
essential quantum effect. 

Since the quantum effect is generally larger for smaller systems, 
the first-order SF transition is more difficult to occur 
in finite clusters, 
treated in the present study, than in the infinite system. 
Therefore the conclusion that SF due to SIA  
can occur for 
$S\geq {3\over 2}$ in 1D is expected to be valid 
even in the thermodynamic limit.
In contrast, 
the absence of SF for $S=1$ should be 
checked by some other methods like the Monte Carlo  
or DMRG\cite{white} to treat larger chains. 
It would be also interesting to perform a magnetization measurement on 
some 1D larger-$S$ materials with negative $D$ ($|D|<J$), 
where the first-order transition is expected 
to occur, and to compare the curve with that of the $S=1$ chain  
CsNiCl$_3$.  

In summary the finite-cluster analysis 
on 1D $S={3\over 2}$ and 2 antiferromagnets suggested 
that the first-order
SF transition can occur for $S\geq {3\over 2}$ in the 
presence of the Ising-like SIA.  
We also discussed on 
all possible magnetization processes for $D\leq 0$  
including the case when two successive second-order transitions appear 
instead of SF. 

It is a pleasure to thank K. Totsuka, H. Nojiri and Y. Ajiro for
helpful discussions.
The numerical computation was done using the facility of the
Supercomputer Center, Institute for Solid State Physics, University of
Tokyo.

\widetext
\end{document}